\documentclass[twocolumn,pra,aps,superscriptaddress,showpacs]{revtex4}
\usepackage{graphicx}
\begin{document}
\title {Solitons and vortices in an evolving Bose-Einstein condensate}
\author{Shi-Jie Yang}
\email{yangshijie@tsinghua.org.cn}
\author{Quan-Sheng Wu}
\author{Shiping Feng}
\affiliation{Department of Physics, Beijing Normal University,
Beijing 100875, China}
\author{Yu-Chuan Wen}
\affiliation{Department of Physics, Capital Normal University,
Beijing 100037, China}
\author{Yue Yu}
\affiliation{Institute of Theoretical Physics, Chinese Academy of
Sciences, P.O. Box 2735, Beijing 100080, China}

\begin{abstract}
Spatiotemporal evolution of a confined Bose-Einstein condensate is
studied by numerically integrating the time-dependent
Gross-Pitaevskii equation. Self-interference between the
successively expanding and reflecting nonlinear matter waves results
in spiral atomic density profile, which subsequently degenerates
into an embedding structure: the inner part preserves memory of the
initial states while the outer part forms a sequence of
necklace-like rings. The phase plot reveals a series of discrete
concentric belts. The large gradients between adjacent belts
indicate that the ring density notches are dark solitons. In the
dynamical process, a scenario of vortex-antivortex pairs are
spontaneously created and annihilated, whereas the total vorticity
keeps invariant.
\end{abstract}
\maketitle

\section{introduction}
Interference experiments demonstrated that the dilute atomic
Bose-Einstein condensate (BEC) is a coherent matter
wave.\cite{Andrews,Wheeler,Hadzibabic} When two spatial separated
phase-independent BECs are left to expand, they overlapp to form
interference fringes. A recent experiment by D.R. Scherer et
al\cite{Scherer} revealed that the interference of three independent
trapped BECs results in the creation of vortices, which gives
definite verification of such nonlinear interference. These
experiments and the relevant theoretical studies opened a new arena
to explore quantum phenomena at the macroscopic level, as well as
nonlinearity effects on the interference and evolution of matter
waves.\cite{Reinhardt,Albiez} The nonlinearity in BECs arises from
the mean-field approximation for short-range interactions between
Bose atoms. It accounts for the existence of many coherent nonlinear
structures that have been observed in experiments, such as
dark,\cite{Burger,Denschlag,Anderson}
bright\cite{Strecker,Khaykovich} and gap solitons,\cite{Eiermann}
vortices\cite{Madison,Inouye,Matthews} and vortex
lattices,\cite{Abo,Engels} etc.

In the BEC, dark soliton (DS) is characterized by a local density
minimum and a sharp phase gradient of the wave function at the
position of the minimum. Burger et al \cite{Burger} carried out an
experiment to create dark solitons in a strongly elongated
condensate by the phase imprinting technique. Theocharis et al
\cite{Theocharis} introduced the concept of ring dark soliton (RDS)
in a 2-dimensional (2D) Bose condensate. It is found that the RDS
solutions are unstable to snake instability, whereby they decay into
vortex-antivortex (V-AV) pairs. The experimentally observed
dynamical instability \cite{Denschlag} is due to their quasi-1D
character, i.e., a DS stripe becomes unstable against transverse
snaking \cite{Brand,Feder,Kevrekidis}.

The tunability of BEC settings allows for rapid change of system
parameters and observation of the subsequent quantum dynamics, which
can remain coherent for exceedingly long times because the
condensate atoms can be well isolated from the environment. It is
also possible to provide means to probe non-equilibrium quantum
dynamics of many body
systems.\cite{Greiner1,Greiner2,Zwierlein,Kollath} In the sudden
change limit one can consider that the system is prepared in the
ground state of an initial Hamiltonian $H_i$, and then evolves under
the influence of a final Hamiltonian $H_f$. For example, in a
one-dimensional optical lattice experiment, the authors found that
when the system parameter of BEC in superfluid regime is suddenly
changed to the deep Mott insulator regime, i.e., $U/t$ large limit,
the BEC may revive to the superfluid regime regularly.

It is well-known that a quantized vortex in superfluid cannot simply
fade away or disappear, it is only allowed to move out of the
condensate or annihilate with another vortex of an opposite
circulation. In almost pure condensates, vortices with lifetimes up
to tens of seconds have been observed.\cite{Matthews,Abo} In this
work we report the dynamical evolution of a BEC confined in a
2-dimensional cylinder well. The initial state is prepared by
loading the condensate into a narrower harmonic well. By using a
topological phase imprinting method we can create one or several
vortices in the condensate. The harmonic well is then suddenly
changed to a wider cylinder well and the BEC is left to freely
expand until it reaches the inner wall of the cylinder well. The
reflecting waves then interfere successively with the expanding
waves. In a previous publication,\cite{yang} we have studied the
evolution of the same setup without vortex in the initial state. We
demonstrated that ring dark solitons are spontaneously generated and
survive for brief time before they are taken place by another set of
solitons. In this paper, we further explore the dynamical evolution
for the Bose condensate with one or several vortices in the initial
state. We find that the condensate first forms a spiral atomic
density profile. Subsequently, the system evolve into a peculiar
embedding pattern with an inner part that seems to preserve the
memory of the initial state. The outer part is a sequence of
necklace-like rings. As in the non-vortex case, The notches between
these density rings are identified as dark solitons, which precess
around the inner kennel. The system evolves with co-existence of
vortices and dark solitons. A scenario of vortex-antivortex creation
and annihilation takes place in the dynamical evolution. But the
total vorticity remains invariant and the density distribution
preserves the initial symmetry.

\section{the model}
In weak interaction limit, quantum dynamics of the BECs are
determined by the nonlinear Gross-Pitaevskii (G-P) equation. We
restrict the problem to the 2D plane and employ a cylindrical trap.
We model the cylindrical well with
$V(\vec{r})=V_0[\tanh(\frac{r^2-r_0^2}{a_0^2})+1]$, where $r_0$ is
the radius of the trap and $a_0$ is a scaling length parameter
relevant to the size of the well. The corresponding characteristic
frequency is $\omega_c=\hbar/ma_0^2$. When $V_0$ is large enough,
the condensate is completely confined. After re-scaling the
parameters by making substitution $t\rightarrow \omega_c t$, $\vec
r\rightarrow \vec r/a_0$, and $\psi\rightarrow \psi/a_0^{3/2}$, one
obtains the reduced dimensionless G-P equation
\begin{equation}
i\frac{\partial \psi}{\partial t}=-\frac{1}{2}\nabla^2
\psi+V(\vec{r}) \psi+ c |\psi|^2 \psi, \label{GP}
\end{equation}
where $c=4\pi N a_s/a_0$ and $a_s$ the s-wave scattering length.

In order to study the quantum dynamics we prepare a steady initial
state by employing a narrower harmonic potential
$V_0(\vec{r})=\alpha r^2/2$, where $\alpha$ is an adjustable
parameter, and let the wavefunction propagate along the imaginary
time of the G-P equation. Quantized vortices are created by using
topological phase imprinting technique\cite{Matthews,Leanhardt}
which is most convenient for rapid preparation of well-defined
vortex states. Subsequently, at $t=0$, the harmonic well is lifted
and the condensate is allowed to freely expand until it reaches the
inner wall of a cylinder well. We investigate the dynamical
evolution of the interfering matter waves. The simulation is carried
out in a region $(x,y)\in [-10,10]\times [10,10]$ with a refined
grid of $256\times 256$ nodes, which is sufficient to achieve grid
independence. Setting the total time scale as $t=15$ and the time
step $\Delta t=0.00005$, we numerically integrate the time-dependent
G-P equation by using the Crank-Nicolson scheme.\cite{Aftalion}

\section{numerical simulations}
Figure 1 snapshots the dynamical evolution of a BEC with one vortex
at the center of the symmetrical well. In the earlier stages, the
condensate freely expands to fill the empty space until it reaches
the wall of the cylinder well. Then the reflecting waves and the
expanding waves interfere to form concentric density rings. The
single vortex always stays at the center. Although vortex is
excitation of motion and therefore energetically unstable towards
relaxation into the ground state which is static, long-range quantum
phase coherence regulates the dynamics of quantized
vortices.\cite{Donnell} The lower panels of Fig.1 display the
corresponding spatial phase distribution of the condensate. As time
evolving, the phase begins to skew to form a helix, indicating that
while expansion the circular motion of the condensate is not
uniform. Novelly, when the evolving time is long enough ($t>5$), the
helix breaks up to form a sequence of discontinuous belts. In the
mean time, an inner kernel mimics the initial state is formed,
surrounded by a sequence of concentric density rings.

In a previous publication, We have identified that the concentric
density notches are ring dark solitons, which survive for a brief
time before evolving into another set of dark solitons. The key
point for the existence of RDS is the large phase gradients across
the density notches. The proof applies for the present case. The
details of the demonstration can be found in Ref.\cite{yang}.

\begin{figure}[t]
\includegraphics[width=8cm]{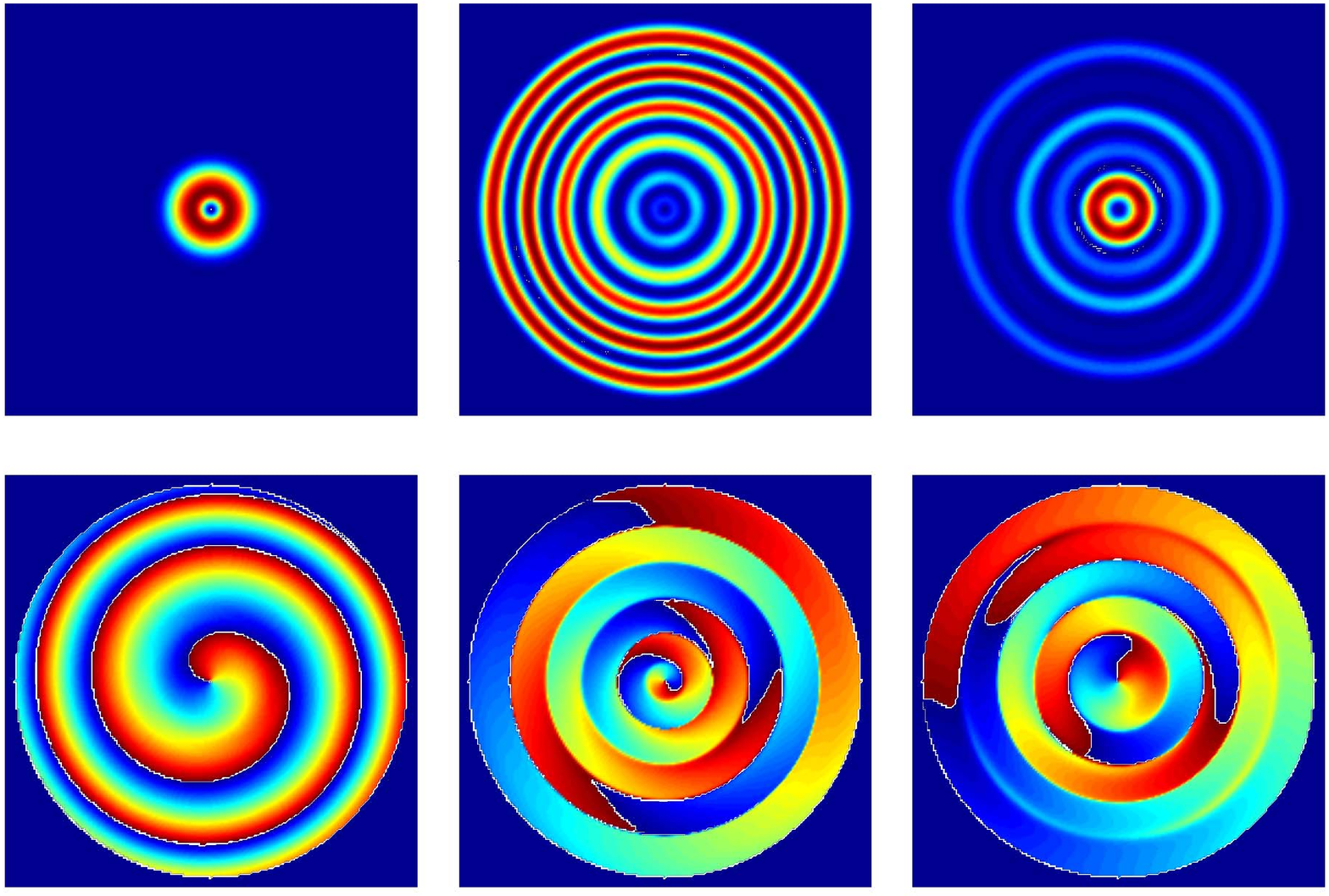}
\caption{(Color online) Snapshots of the density (upper panels) and
phase distribution (lower panels) in a region $\{X,Y\}\in [-10,10]$
(in units of $a_0$). In the initial state there is a vortex at the
origin. From left to right: $t=0.38$, $t=5$ and $t=10$ (in units of
$\omega_c^{-1}$). As $t\gtrsim 5$, the phase rings gradually
degenerate into discontinuous belts.}
\end{figure}

Next we investigate the dynamical evolution of the condensate with
four symmetric vortices in the initial state. In this case, the
system has a four-fold rotational symmetry. The upper panels of
Fig.2 shows the evolution of the BEC density profile at three
typical moments $t=0.38$, $5$, and $10$. When the condensate is
released from the initial potential, the four vortices expand
outward and gradually disappear due to self-interference. At the
same time, four vortices somehow reappear around the central area.
For this somehow more complex initial state, the density forms a
four-fold symmetrical spiral structure. However, these quasi-1D
spiral density arms are unstable towards transverse modulations. As
a result, the density arms gradually break up and a peculiar
embedding structure forms when $t\gtrsim 6$ (see left panel in
Fig.2). The initial four vortices muster at the central area,
surrounded by a sequence of necklace-like density rings. The inner
part is quite similar to the initial state. The outer part is again
the rotated necklace-like dark solitons. After this evolving
process, the condensate seems to \textit{throw off} part of its mass
to fill the empty space and form ring dark solitons, leaving an
inner kernel that preserves the initial momery.

\begin{figure}[t]
\includegraphics[width=8cm]{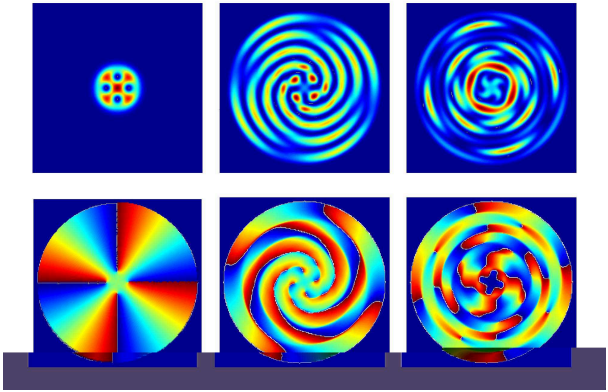}
\caption{(Color online) Snapshots of the density (upper panels) and
phase distribution (lower panels) at $t=0.01$, $t=5$ and $t=10$. In
the initial state there are four symmetric vortices at the center.}
\end{figure}

The lower panels of Fig.2 show the corresponding phase plot of the
evolving condensate. At the earlier stage, the phase skews into
helices. This is because the condensate has nonzero total angular
momentum and rotates around the symmetrical axis. But when the
evolving time is long enough ($t\gtrsim 6$), the phase plot again
shows a sequence of plateau-like belt. Abrupt jumps between adjacent
phase belts indicate the formation of solitons. We note that the
dynamical soliton can last for a brief period but is still evolving.
As estimated in Ref.\cite{yang}, the duration of the solitons is
about $0.5$ms for $c=20$.

The instability of the spiral stripes results in rich phenomena of
creation and annihilation of V-AV pairs. Figure 3 displays a typical
spatial superflow distribution at $t=5$. The evolution of V-AV pairs
in the present BEC setting is comparable to the previous
work,\cite{Feder,Kevrekidis,Whitaker} where several separated BECs
were allowed to expand and merge together. Although the creation and
annihilation of the V-AV pairs is rather complex, the whole
distribution keeps a perfect four-fold symmetry. After a long time,
this V-AV scenario gradually fades away when the density profile
wears the concentric necklace-like solitons.

\begin{figure}[t]
\includegraphics[width=8cm]{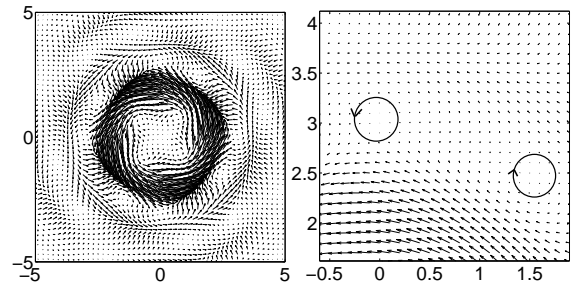}
\caption{Vectorial plot of the super-flow with four vortices at
$t=5$. The right panel is an enlarged local regime. The arrowed
circles indicate the positions of a vortex and an antivortex.}
\end{figure}

Figure 4 takes snapshots of the density and phase distribution with
seven vortices in a six-fold symmetric initial state, in which one
vortex locates at the origin surrounded by the other six vortices.
Analogous to the case of four vortices, six spiral density arms are
first derived from the condensate. Then these arms are broken by
continuous interference and oscillations. An embedding structure
again appears. The inner kernel resembles the initial state. In the
course of dynamical evolution, the density distribution always
exhibits a six-fold symmetry and the total vorticity keeps
invariant.

\begin{figure}[t]
\includegraphics[width=8cm]{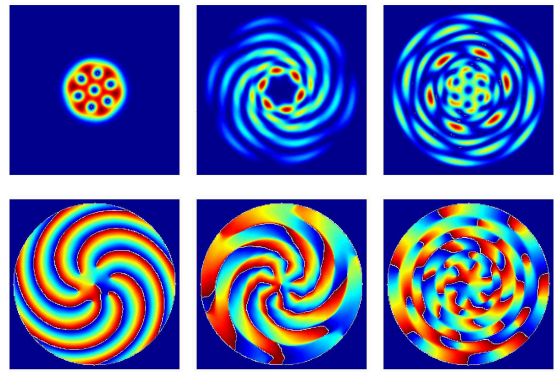}
\caption{(Color online) Same as in Fig.1 for BEC with seven
symmetric vortices at the center. The snapshot times from left to
right are $t=0.38$, $t=5$ and $t=10$.}
\end{figure}

\section{summary}
We have studied the dynamical evolution of a repulsive Bose-Einstein
condensate in a 2D external potential. We found that the density
profile shows an embedding structure in which the inner kernel
resemble the initial state while the outer part forms a sequence of
necklace-like solitons, which circulates around the inner kennel.
These solitons survives for a brief time before evolving into
another set of solitons. A scenario of creation and annihilation of
vortex-antivortex pairs takes place in the evolving process.

{\bf Acknowledgement} This work is supported by the National Natural
Science Foundation of China under grant No. 10574012.

\end{document}